# Bounded Rationality, Strategy Simplification, and Equilibrium

Hubie Chen
Departament de Tecnologies de la Informació i les Comunicacions
Universitat Pompeu Fabra
Barcelona, Spain
hubie.chen@upf.edu

## Abstract

It is frequently suggested that predictions made by game theory could be improved by considering computational restrictions when modeling agents. Under the supposition that players in a game may desire to balance maximization of payoff with minimization of strategy complexity, Rubinstein and co-authors studied forms of Nash equilibrium where strategies are maximally simplified in that no strategy can be further simplified without sacrificing payoff. Inspired by this line of work, we introduce a notion of equilibrium whereby strategies are also maximally simplified, but with respect to a simplification procedure that is more careful in that a player will not simplify if the simplification incents other players to deviate. We study such equilibria in two-player machine games in which players choose finite automata that succinctly represent strategies for repeated games; in this context, we present techniques for establishing that an outcome is at equilibrium and present results on the structure of equilibria.

## 1 Introduction

A frequently raised criticism of game theory is that its predictions clash with empirical observations as a consequence of being based on the assumption that agents possess and use unbounded computational power. This criticism has motivated the introduction and investigation of so-called models of *bounded rationality* [28], in which computational power considerations on agents are present. A number of different approaches to the study of bounded rationality have been suggested [2, 7, 17, 24, 26, 27]. One model that has received considerable attention is a machine game in which players choose finite-state automata that succinctly represent strategies for repeated games [18, 21, 24, 26, 1]; the model of finite-state automata can be taken as a formalization of players having bounded-size memory, and is well-studied in computer science.

Rubinstein with co-authors Abreu and Piccione [26, 1, 25], in the context of the machine game, proposed and studied forms of Nash equilibrium under which strategies are maximally simplified in the sense that a player's strategy cannot be simplified without reducing his payoff. A supposition basic to this work is that players desire to minimize the complexity of their strategies, and hence in choosing strategies are concerned with balancing the maximization of payoff with the minimization of strategy complexity. Simplicity of strategies may be valued for a number of reasons; for instance, complex strategies may be more expensive to execute, more likely to break down, harder to learn, or costly to maintain [22]. Following Rubinstein [26], it can be suggested that such maximally simplified equilibria resemble phenomena observed in real life:

institutions, organizations, and human abilities may degenerate or be reduced if they contain unnecessary or redundant components.

The Rubinstein-Abreu-Piccione line of work, more specifically, studies Nash equilibria where the players have preference relations that increase in the payoff and, when payoff is maintained, decrease in the complexity. While one can study the equilibria they define in any game where there is a complexity measure associated with the actions of each player, their work focuses on the mentioned machine game, and the complexity measure studied is the number of states of an automaton, which can be viewed as the memory size of a strategy.

Inspired by this line of work, we introduce and study a new notion of equilibrium intended to capture maximally simplified strategies, but with respect to a more careful, conservative simplification procedure. The motivation for this new equilibrium notion stems from the observations that, in the Rubinstein model, a player simplifies without considering whether or not the simplification may incent other players to deviate, and that this liberal mode of simplification may spoil desirable outcomes that are Nash equilibria in the usual payoff sense. These observations can be illustrated by the following example. Consider the so-called grim trigger strategy in the infinitely repeated Prisoner's dilemma; this strategy, which can be implemented by a two-state automaton, is to cooperate until the other player is seen to have defected, and to then defect indefinitely. While this strategy paired with itself is a Nash equilibrium in the payoff sense–no player can unilaterally deviate and increase his payoff–it is not an equilibrium in the Rubinstein sense, since either player could maintain payoff but reduce complexity by switching to a strategy that always cooperates (which is a strategy that can be implemented by a one-state automaton). Notice, however, that such a switch would in turn incent the other player to change to a strategy that always defects against cooperation, thus spoiling the cooperation. It can in fact be verified that no pair of strategies that cooperate indefinitely form an equilibrium in the Rubinstein sense.

Whereas in the Rubinstein model a player will simplify his strategy so long as he can maintain his payoff, in our model each player is forward-looking, and will only simplify his strategy if, in addition, no other player can profitably deviate post-simplification. That is, in considering simplifications, players are averse to potential payoff-motivated deviations by other players. Our notion of equilibrium, which we call *lean equilibrium*, is thus defined as an outcome of strategies at Nash equilibrium such that no player can both individually simplify his strategy *and* preserve the property of being at Nash equilibrium. The described grim trigger strategy paired with itself does constitute a lean equilibrium in the infinitely repeated prisoner's dilemma: the described strategy has two states, so any simplification must have one state; in order for the result to be a Nash equilibrium, the player with one state must always cooperate in order to be a best response to the other player; but, this is not a Nash equilibrium as the other player could then profitably deviate by always defecting.

We present results on lean equilibria for two-player machine games where each player chooses a finite-state automaton representing a strategy in an infinitely repated game. We study three complexity measures; in addition to studying the "number of states" measure, we study two measures that we introduce. One is based on the number of states, but does not count threat states, and the other counts the number of transitions to non-threat states; the precise definitions appear later in the paper. Our primary technical results are the following.

- We give techniques for establishing that outcomes are at lean equilibrium, and illustrate their use by a number of examples (Section 5).

- We present results on the structure of machines that are at equilibria, and, with respect to the number-of-transitions measure, give a precise description of the equilibria structure. This description in fact shows that the machine structure can be inferred from a third-party observer that only views the induced sequence of action pairs (Section 6).

We believe that the developed theory evidences that the two introduced complexity measures are natural and mathematically robust.

While the present work focuses on machine games with finite automata and was certainly inspired by previous work on such games, we want to emphasize that the notion of lean equilibrium is defined in a very general way (Section 3) and can be applied to any game in which there is a notion of complexity associated with the players' actions. Indeed, our view is that one of the most promising avenues for future work is to analyze the lean equilibria in other types of games where such a notion of complexity is present or can be naturally defined; in particular, it could be of interest to study games arising directly from real-life situations and phenomena. We believe that the theory and results developed in this work vindicates the introduced equilibrium notion as a tangible and robust mathematical concept of which one can hope to present analysis in further games.

**Related work.** The present work is a contribution to the study of bounded rationality; surveys and general references include [27, 17, 3]. The present article can in particular be taken as following a body of research where players in games are represented using models of computation; here, we briefly describe some of this research.

The study of machine games where players select automata representing strategies in repeated games was initiated early in the study of bounded rationality. The already described work of Rubinstein, Abreu, and Piccione studied Nash equilibria where players' preferences take into account strategy complexity in addition to repeated game payoff. The paper of Rubinstein [26] studied an equilibrium concept in the spirit of subgame perfect equilibrium, obtaining structural results on such equilibria. Abreu and Rubinstein [1] gave general structural results on equilibria and studied the payoff sets of certain 2-by-2 games; and, Piccione and Rubinstein [25] studied equilibria in repeated extensive games. Banks and Sundaram [4] studied a equilibrium notion similar to that considered in these papers, but focused on a "transitional" notion of strategy complexity that accounts for the amount of opponent monitoring required, and can differentiate among strategies with the same number of states. Kalai and Stanford [18] gave a characterization of the number-of-states complexity measure for automata via an analog of the Myhill-Nerode theorem, and study subgame perfect equilibria in infinitely repeated games from the viewpoint of this measure. A number of works studied repeated games played by finite automata where bounds on the strategy complexity are imposed exogenously, including the articles of Neyman [20], Ben-Porath [6], Papadimitriou and Yannakakis [24], and Neyman [21]; one focus of study is the set of payoffs sustainable in equilibrium. Gilboa [13], Papadimitriou [23], and Ben-Porath [5] studied the computational complexity of problems involving the computation of best response automata. Spiegler [29, 30] presented equilibrium notions motivated by the idea that players may need to justify their strategies; he modeled players as finite-state automata.

Another line of work studies games where players must employ computable strategies. Some of the initial work explored basic consequences of this modeling and invokes notions and ideas from computability theory, including the contributions of Binmore [8, 9], Canning [10], and Anderlini [2]. Megiddo and Wigderson [19] presented results on games played by Turing machines where the number of states is restricted. Howard [16], Tennenholtz [31], and Fortnow [11] showed existence of equilibrium and folk theorem style results by making use of self-reference ideas. Gossner [14] studied repeated games played by polynomial-time Turing machines, invoking cryptographic assumptions to obtain results on the equilibria achievable under public communication.

More recent work includes the following. A framework for games where the actions have associated costs was proposed and studied by Ben-Sasson, Kalai, and Kalai [7]. Halpern and Pass [15] presented and studied a machine game on Turing machines where utilities can be a function of machine complexities in addition to the action profile; among other issues, they study existence of equilibria and notions of protocol security. Fortnow and Santhanam [12] introduced and studied a machine game model where players' actions are probabilistic Turing machines that output actions in an underlying game; the payoff associated with a machine is discounted by the computation time used to produce actions. Their results include connections between the existence of Nash equilibria in the so-called factoring game and the computatoinal complexity of factoring, and general sufficient conditions for the existence of equilibria.

## 2 Preliminaries

In this section, we review some basic notions to be used. Our notation and terminology are standard, and for the most part follow typical conventions such as those described in the text by Osborne and Rubinstein [22]. A *strategic game* is a tuple $(N, (A_i), (\preceq_i))$ consisting of a set $N = \{1, \ldots, n\}$ of players, a nonempty set $A_i$ of actions for each player $i \in N$, and a preference relation $\preceq_i$ defined on $A = \times_{j \in N} A_j$ for each player $i \in N$. For the most part, we will focus on two-player strategic games where the preference relations are specified by payoff functions $u_i : A_i \to \mathbb{R}$. Recall that a *Nash equilibrium* of a strategic game $(N, (A_i), (\preceq_i))$ is a profile $a^* \in A$ of actions such that for all $i \in N$ and for all $a_i \in A_i$, it holds that $(a^*_{-i}, a_i) \preceq_i (a^*_{-i}, a^*_i)$.

For our purposes in this paper, a *convex combination* of vectors $x_1, \ldots, x_d \in \mathbb{R}^m$ is a vector of the form $\alpha_1 x_1 + \cdots + \alpha_d x_d$ where the $\alpha_n$ are *rational* coefficients with $\sum_{n=1}^d \alpha_n = 1$ and $0 \leq \alpha_n \leq 1$ for each $n$.

We will make use of the following payoff notions. Let $G = (N, (A_i), (u_i))$ be a strategic game. A *feasible payoff profile* of $G$ is a convex combination of the vectors $\{u(a) \mid a \in A\}$. We define player $i$'s *minmax payoff*, denoted $v_i$, to be the lowest payoff that the other players can force upon player $i$, that is, $v_i = \min_{a_{-i} \in A_{-i}} \max_{a_i \in A_i} u_i(a_{-i}, a_i)$. A feasible payoff profile $w \in \mathbb{R}^n$ of $G$ is called *enforceable* if $v_i \leq w_i$ for all $i \in N$, and is called *strictly enforceable* if $v_i < w_i$ for all $i \in N$. See Osborne and Rubinstein [22, Section 8.5] for more information on these notions.

## 3 Lean equilibrium

We define a *complexity order* on a set of actions $A_i$ to be a binary relation on $A_i$. We will consider games where each player $i \in N$ has a complexity order $\trianglelefteq_i$ associated to his set of actions $A_i$; the intended interpretation is that $b_i \trianglelefteq_i a_i$ if player $i$ considers the action $b_i$ to have the same complexity as or lower complexity than action $a_i$. In such games, for $a_i, b_i \in A_i$, we will write $b_i \triangleleft a_i$ to denote that $b_i \trianglelefteq a_i$ holds and $a_i \trianglelefteq b_i$ does not hold. Also, for $a, b \in A$, we will write $b \trianglelefteq a$ to denote that for all $i \in N$, it holds that $b_i \trianglelefteq_i a_i$. We remark that in studying machine games, each complexity order that we consider arises from associating machines with elements in a total order; however, for broadest applicability, the results in this section (in particular, Proposition 3.2) are presented for more general settings.

**Definition 3.1** Let $G = (N, (A_i), (\preceq_i))$ be a strategic game with complexity orders $(\trianglelefteq_i)$. A profile $a^* \in A$ of actions is a *lean equilibrium* of the game $G$ if $a^*$ is a Nash equilibrium, but for all $i \in N$ and for all $a_i \in A_i$, if $a_i \triangleleft a^*_i$, then $(a^*_{-i}, a_i)$ is not a Nash equilibrium. □

We now present a basic property of lean equilibrium, namely, the existence of lean equilibria under the assumption of the existence of Nash equilibria and a mild assumption on the complexity orders.

**Proposition 3.2** *Suppose that $G = (N, (A_i), (\preceq_i))$ is a strategic game with complexity orders $(\trianglelefteq_i)$ that are transitive and are well-founded in the sense that for all $i \in N$ and for all $a_i \in A_i$, there exists a bound on the length of a chain $c_1 \triangleleft_i \cdots \triangleleft_i c_k \triangleleft_i a_i$. Then for every Nash equilibrium $a^*$ of G, there exists a lean equilibrium $b \in A$ such that $b \trianglelefteq a^*$.*

**Proof.** For an action $a_i$ from an action set $A_i$, let $C(a_i)$ denote the maximum length $k$ of a chain $c_1 \triangleleft_i \cdots \triangleleft_i c_k \triangleleft_i a_i$. For an action profile $a \in A$, define $C(a) = \sum_{i \in N} C(a_i)$. We prove the result by induction on $C(a^*)$.

For the base case, where $C(a^*) = 0$, the profile $a^*$ is a lean equilibrium, as for all $i \in N$ and for all $a_i \in A_i$, it does not hold that $a_i \triangleleft_i a_i^*$. For the inductive case, suppose that $C(a^*) > 0$. If the profile $a^*$ is not a lean equilibrium, then there exists $i \in N$ and there exists $a_i \in A_i$ such that $(a_{-i}^*, a_i)$ is a Nash equilibrium and $a_i \triangleleft_i a_i^*$. We have that $C(a_{-i}^*, a_i) < C(a^*)$; by applying the induction hypothesis to $(a_{-i}^*, a_i)$, we obtain a lean equilibrium $b \in A$ such that $b \trianglelefteq (a_{-i}^*, a_i)$. We have $(a_{-i}^*, a_i) \trianglelefteq a^*$ and thus by transitivity of $\trianglelefteq$, it holds that $b \trianglelefteq a^*$. □

One of the equilibrium notions studied by Abreu and Rubinstein [1] is Nash equilibrium with respect to the lexicographical ordering where payoff is prioritized over complexity: one profile $a^*$ is strictly preferred by a player to another profile $b^*$ if $a^*$ gives the player a strictly higher payoff than $b^*$, or if $a^*$ gives the player the same payoff as $b^*$ but the player has strictly lower complexity in $a^*$. We formalize this equilibrium notion and show that each such equilibrium is a lean equilibrium, as follows.

**Definition 3.3** *Let $G = (N, (A_i), (\preceq_i))$ be a strategic game with complexity orders $(\trianglelefteq_i)$. A profile $a^* \in A$ of actions is an* Abreu-Rubinstein equilibrium *of the game G if for all $i \in N$ and for all $a_i \in A_i$, (1) $(a_{-i}^*, a_i) \preceq_i (a_{-i}^*, a_i^*)$ and (2) $(a_{-i}^*, a_i^*) \preceq_i (a_{-i}^*, a_i)$ implies that $a_i \triangleleft_i a_i^*$ does not hold.* □

Condition (1) implies that the profile $a^*$ is a Nash equilibrium, and condition (2) essentially says that there is no deviation $a_i$ for player $i$ that yields the same utility and is simpler than $a_i^*$.

**Proposition 3.4** *Let $G = (N, (A_i), (\preceq_i))$ be a strategic game with complexity orders $(\trianglelefteq_i)$. Every Abreu-Rubinstein equilibrium is a lean equilibrium.*

**Proof.** Let $a^*$ be an Abreu-Rubinstein equilibrium. Clearly, the profile $a^*$ is a Nash equilibrium. Now consider a player $i \in N$ and an action $a_i \in A_i$. We want to show that if $a_i \triangleleft a_i^*$, then $(a_{-i}^*, a_i)$ is not a Nash equilibrium. By the definition of Abreu-Rubinstein equilibrium, if $a_i \triangleleft a_i^*$, then $(a_{-i}^*, a_i) \prec_i (a_{-i}^*, a_i^*)$, implying, as desired, that $(a_{-i}^*, a_i)$ is not a Nash equilibrium. □

In light of this proposition, all examples of Abreu-Rubinstein equilibria are examples of lean equilibria. For instance, the Abreu-Rubinstein equilibria of certain two-player games are studied in [1, Section 5]; all examples given there are examples of lean equilibria. On the other hand, later in this paper, we will encounter examples of lean equilibria that are not Abreu-Rubinstein equilibria (for instance, in Examples 5.5 and 5.9).

## 4 Machine games

In this section, we introduce the machine games whose lean equilibria we will study, and some associated notions. These games involve choosing machines which implement strategies for repeated games, and have been previously studied, as discussed in the introduction. For more background on strategies as machines and for some simple examples, we refer the reader to Osborne and Rubinstein [22, Section 8.4 and Chapter 9].

**Machines and machine games.** Let $G = (S_1, S_2, u_1, u_2)$ be a two-person game in strategic form, where $S_i$ is a finite set of actions for player $i$, and $u_i : S_1 \times S_2 \to \mathbb{R}$ is the payoff function for player $i$.

A *machine* for player $i$ is a four tuple $M_i = (Q_i, q_i^1, \lambda_i, \delta_i)$ where $Q_i$ is a finite set of *states*, $q_i^1 \in Q_i$ is the *start state* or *initial state*, $\lambda_i : Q_i \to S_i$ is the *output function*, and $\delta_i : Q_i \times S_j \to Q_i$ is the *transition function*; here, $S_j$ denotes the action set of the other player. We emphasize that in this paper, we consider only machines that conform to this definition, namely, machines which have a finite number of states and which are deterministic. We use $\mathcal{M}_i$ to denote the set of all machines for player $i$, relative to a game $G$. We often use $i$ and $j$ to denote the two different players.

A pair of machines $(M_1, M_2)$ naturally induces a sequence $(q^t)_{t \geq 1}$ of state pairs and a sequence $(s^t)_{t \geq 1}$ of action pairs ($G$-outcomes), defined inductively as follows:

$$\begin{aligned} q^1 &= (q_1^1, q_2^1) \\ s^t &= (\lambda_1(q_1^t), \lambda_2(q_2^t)) && \text{for } t \geq 1 \\ q^t &= (\delta_1(q_1^{t-1}, s_2^{t-1}), \delta_2(q_2^{t-1}, s_1^{t-1})) && \text{for } t > 1 \end{aligned}$$

Each of the sequences $(q^t)$, $(s^t)$ is *ultimately periodic*; we say that a sequence $(b^t)_{t \geq 1}$ is *ultimately periodic* if there exist numbers $n, p \geq 1$ such that for all $m \geq n$, it holds that $b^m = b^{m+p}$.

The payoff given to each machine is computed by the limit of means. For a sequence of action pairs $(s^t)$, we define $r_i^T(s^t)$ to be the average payoff to player $i$ over the first $T$ elements of the sequence, that is, we define $r_i^T(s^t) = \frac{1}{T} \sum_{k=1}^T u_i(s^k)$. We define $r_i(s^t)$ to be the corresponding limit of the average payoffs, that is, $r_i(s^t) = \lim_{T \to \infty} r_i^T(s^t)$; note that we will make use of this function only on ultimately periodic sequences $(s^t)$, and so the limit will always exist. For a pair of machines $(M_1, M_2)$, we define $r_i^T(M_1, M_2) = r_i^T(s^t)$ and $r_i(M_1, M_2) = r_i(s^t)$, where here $(s^t)$ denotes the sequence of action pairs induced by the machines $(M_1, M_2)$. Our focus will be on the machine game defined by $G_m = (\mathcal{M}_1, \mathcal{M}_2, r_1, r_2)$.

**Paths and cycles.** With respect to a two-player game, a *path* in a machine $M_i$ is a sequence $p_1 \xrightarrow{a_1} p_2 \xrightarrow{a_2} \cdots \xrightarrow{a_m} p_{m+1}$ where $p_1, \ldots, p_{m+1} \in Q_i$ are states, $a_1, \ldots, a_m \in S_j$ are actions, and $\delta_i(p_k, a_k) = p_{k+1}$ for each $k \in \{1, \ldots, m\}$. A *cycle* is a path where $p_1 = p_{m+1}$, and is said to be a *simple cycle* if the states $p_1, \ldots, p_m$ are pairwise distinct. For a path $P$ in $M_1$, the payoff to $M_1$, denoted by $r_1(P)$, is defined as $\frac{1}{m} \sum_{k=1}^m u_1(\lambda_1(p_k), a_k)$; and, the payoff to $M_2$, denoted by $r_2(P)$, is defined as $\frac{1}{m} \sum_{k=1}^m u_2(\lambda_1(p_k), a_k)$. For a path $P$ in $M_2$, the payoffs $r_1(P)$ and $r_2(P)$ are defined similarly. It is known and straightforward to verify that, in the machine game $G_m$, a payoff-maximizing response for player $i$ to a machine $M_j$ has payoff equal to the maximum of $r_i(C)$ over all cycles in the machine $M_j$ reachable from the initial state, which is equal to the maximum of $r_i(C)$ over all such cycles in the machine $M_j$ that are simple.

Let us define a *subcycle* of a cycle $p_1 \xrightarrow{a_1} p_2 \xrightarrow{a_2} \cdots \xrightarrow{a_m} p_{m+1}$ to be a cycle that has either the form $p_n \xrightarrow{a_n} p_{n+1} \xrightarrow{a_{n+1}} \cdots p_{n'}$ or the form $p_{n'} \xrightarrow{a_{n'}} p_{n'+1} \xrightarrow{a_{n'+1}} \cdots \xrightarrow{a_m} p_{m+1} = p_1 \xrightarrow{a_1} p_2 \xrightarrow{a_2} \cdots p_n$ with $n, n'$ satisfying $1 \leq n \leq n' \leq m$.

**Complexity measures.** We will study three complexity measures on machines. The first is the number of states $|Q_i|$ of a machine $M_i$. We define the other two in the following way. We define a *threat state* of a machine $M_i$ to be a state $q$ such that $\delta_i(q, s) = q$ for all $s \in S_j$ and where $\max_{a_j \in A_j} u_j(\lambda_i(q), a_j)$ is the minmax payoff of the other player $j$, that is, where $\lambda_i(q)$ forces the other player $j$ to his minmax payoff. We define a *normal state* of a machine $M_i$ to be a state that is not a threat state. We use $R_i$ to denote the set of all normal states of a machine $M_i$, and we use $||\delta_i||$ to denote the number of *normal transitions*, by which we mean transitions between normal states:

$$||\delta_i|| = |\{(q_i, s_j) \in R_i \times S_j \mid \delta_i(q_i, s_j) \in R_i\}|.$$

The two other complexity measures that we will study are the number of normal states, denoted by $|R|$, and the number of normal transitions, denoted by $||\delta||$. We will speak of lean equilibria with respect to, for instance, the measure $|R|$, by which we mean a lean equilibria where the complexity order for player $i$ is given by $M_i \trianglelefteq_i M_i'$ if and only if $|R_i| \leq |R_i'|$; here, $R_i$ and $R_i'$ denote the sets of normal states of the machines $M_i$ and $M_i'$, respectively.

Relative to a pair of machines $(M_1, M_2)$, for each $i \in \{1, 2\}$, we define the set of *played states* for player $i$, denoted by $P_i$, to be the set $\{q_i^t \mid t \geq 1\}$; here, $(q^t)$ denotes the sequence of state pairs induced by $(M_1, M_2)$. We identify the following facts concerning played states, which we will sometimes use tacitly in the sequel.

**Proposition 4.1** *Let $(M_1, M_2)$ be a pair of machines having a strictly enforceable payoff profile. For each player $i \in \{1, 2\}$:*

- *Every played state is a normal state, and thus it holds that $|P_i| \leq |R_i|$.*

- *The number of played states lower bounds the number of normal transitions: $|P_i| \leq ||\delta_i||$.*

The first claim follows from the assumption that the payoff profile is strictly enforceable, which implies that neither player ever plays a threat state. The second claim follows from the first and the observation that every played state has at least one transition to another played state.

**Equivalence relations.** We now introduce a number of equivalence relations, each of which is defined over the set of positive integers, that will be used in our analysis and description of lean equilibria. Let $(s^t)$ and $(q^t)$ be the sequences of action pairs and state pairs, respectively, induced by a pair of machines. We define the equivalence relation $\equiv_s$ by: $t \equiv_s t'$ if and only if for all $n \geq 0$, it holds that $s^{t+n} = s^{t'+n}$. Similarly, we define the equivalence relation $\equiv_q$ by: $t \equiv_q t'$ if and only if for all $n \geq 0$, it holds that $q^{t+n} = q^{t'+n}$. However, from the determinism of the machines, it is straightforward to verify that $t \equiv_q t'$ if and only if $q^t = q^{t'}$; we will make use of this simpler characterization. As the sequence $(s^t)$ is equal to the sequence $(q^t)$ mapped under the functions $(\lambda_1, \lambda_2)$, it is clear that if $t \equiv_q t'$, then $t \equiv_s t'$; viewing these equivalence relations as sets of pairs, we can write $\equiv_q \subseteq \equiv_s$. We define the equivalence relations $\equiv_i$ for $i \in \{1, 2\}$ by $t \equiv_i t'$ if and only if $q_i^t = q_i^{t'}$. It is clear that $t \equiv_q t'$ if and only if $t \equiv_1 t'$ and $t \equiv_2 t'$. For a value $t \geq 1$, we will use $[t]_s$ to denote the $\equiv_s$-equivalence class of $t$, and similarly for the other equivalence relations.

## 5 Establishing lean equilibrium: examples and theory

In this section, we give techniques for establishing that outcomes of the machine game are at lean equilibrium, and illustrate their use by presenting a number of examples. We begin by defining some notions; the definitions and also the later results are relative to a game $G = (S_1, S_2, u_1, u_2)$ and its corresponding machine game $G_m$, although in what follows we will generally not mention the games $G$ and $G_m$ explicitly.

By a *finite action sequence*, we mean a finite-length sequence $\sigma = \sigma^1 \ldots \sigma^k$ of action pairs (elements of $S_1 \times S_2$). For each player $i \in \{1, 2\}$, we define the payoff of a finite action sequence $\sigma$ as $r_i(\sigma) =$ $(u_i(\sigma^1) + \cdots + u_i(\sigma^k))/k$. We say that a finite action sequence $\sigma$ is *strictly enforceable* if its payoff profile $(r_1(\sigma), r_2(\sigma))$ is strictly enforceable. Each strictly enforceable finite action sequence $\sigma$ naturally induces a pair of machines $(M_1^\sigma, M_2^\sigma)$, where for each player $i \in \{1, 2\}$, the machine $M_i^\sigma$ is defined to have $k+1$ states: $k$ normal states, which we denote as $\{1, \ldots, k\}$, and a threat state. The output function $\lambda_i$ is defined with $\lambda_i(n) = \sigma_i^n$ for all $n \in \{1, \ldots, k\}$. The transition function $\delta_i$ has $\delta_i(n, \sigma_j^n) = n+1$ for $n \in \{1, \ldots, k-1\}$, and $\delta_i(k, \sigma_j^k) = 1$; all other transitions out of the normal states go to the threat state. We use $\langle \sigma \rangle$ to denote the infinite sequence obtained by repeating $\sigma$, that is, the sequence $\sigma\sigma\ldots = \sigma_1 \ldots \sigma_k \sigma_1 \ldots \sigma_k \ldots$. The sequence of action pairs generated by $(M_1^\sigma, M_2^\sigma)$ is clearly equal to $\langle \sigma \rangle$. This property is a motivation for the definition of the machines: the transitions of the machines are designed so that the machines together will generate the sequence $\langle \sigma \rangle$, but each machine will "punish" any deviation from this sequence by moving to and settling upon its threat state. We call a machine $M$ a $\sigma$-*machine* if for any best response $M'$ to $M$, the pair $(M, M')$ generates the sequence $\langle \sigma \rangle$; clearly, for any strictly enforceable sequence $\sigma$, the machines $M_1^\sigma$ and $M_2^\sigma$ are $\sigma$-machines.

We now present the first concepts and results that will allow us to give examples of lean equilibria. Relative to a sequence $(s^t)$ of action pairs we say that two time points $t_1, t_2 \geq 1$ are $i$-*incompatible* if there exists $m \geq 0$ such that (1) for all $n$ with $0 \leq n < m$, it holds that $s_j^{t_1+n} = s_j^{t_2+n}$, and (2) $s_i^{t_1+m} \neq s_i^{t_2+m}$. We say that two equivalence classes $T_1, T_2$ of $\equiv_s$ are $i$-*incompatible* if there exist $t_1 \in T_1$ and $t_2 \in T_2$ such that $t_1$ and $t_2$ are $i$-incompatible; observe that, in fact, if equivalence classes $T_1$ and $T_2$ are $i$-incompatible, then for all $t_1 \in T_1, t_2 \in T_2$ it holds that $t_1$ and $t_2$ are $i$-incompatible.

**Proposition 5.1** *Let $(q^t)$, $(s^t)$ be the state sequence and action sequence induced by a pair of machines. If two $\equiv_s$-equivalence classes $T_1, T_2$ are $i$-incompatible, then for all $t_1 \in T_1$, $t_2 \in T_2$, it holds that $q_i^{t_1} \neq q_i^{t_2}$.*

This proposition follows immediately from the definitions of $(q^t)$ and $(s^t)$.

We say that a finite action sequence $\sigma$ is $i$-*irreducible* if for any two distinct values $t_1, t_2 \in \{1, \ldots, k\}$, it holds that $[t_1]_s$ and $[t_2]_s$ are $i$-incompatible with respect to $(s^t) = \langle \sigma \rangle$.

**Theorem 5.2** *Let $\sigma$ be a strictly enforceable finite action sequence of length $k$, and let $M_j$ be a $\sigma$-machine. If $\sigma$ is $i$-irreducible, then then there are $k \equiv_s$-equivalence classes, $[1]_s, \ldots, [k]_s$; and, for any best response $N_i$ to $M_j$ with $|P_i| \leq k$, the pair $(N_i, M_j)$ has $\equiv_i$ equal to $\equiv_s$.*

**Proof**. By the definition of $i$-irreducibility and the definition of $\equiv_s$, we have that there are $k \equiv_s$-equivalence classes, $[1]_s, \ldots, [k]_s$. Consider a best response $N_i$ to $M_j$. Since $M_j$ is a $\sigma$-machine, the pair $(N_i, M_j)$ must produce an action sequence $(s^t)$ equal to $\langle \sigma \rangle$. By Proposition 5.1, no $i$-state is played in two different $\equiv_s$-equivalence classes. Since by hypothesis the number of states played by $i$ is less than or equal to $k$, we have that $[1]_s, \ldots, [k]_s$ must be the equivalence classes of $\equiv_i$, and hence that $\equiv_i$ is equal to $\equiv_s$. □

In this section, we will give a number of examples involving the Prisoner's Dilemma, which we take to be the following game:

|   | C | D |
|---|---|---|
| C | (2, 2) | (-1, 3) |
| D | (3, -1) | (0, 0) |

Note that, in this game, each of the two players has a minmax payoff of 0. For an integer $N \geq 0$ and an action pair $(s_1, s_2)$, we will use the notation $N \cdot (s_1, s_2)$ to denote the finite action sequence containing $N$ copies of the pair $(s_1, s_2)$. For instance, $2 \cdot (C, D)$ represents the sequence $(C, D), (C, D)$ and $2 \cdot (D, C), 3 \cdot (C, D)$ represents the sequence $(D, C), (D, C), (C, D), (C, D), (C, D)$.

**Example 5.3** Let $N_C, N_D \geq 1$ be constants, and consider the finite action sequence $\sigma = N_C \cdot (C,C), N_D \cdot (D,D)$ in the Prisoner's Dilemma. Clearly, the sequence $\sigma$ is strictly enforceable, and has length $k = N_C + N_D$. We show that, with respect to the measures $|R|$ and $||\delta||$, the pair $(M_1^\sigma, M_2^\sigma)$ is an Abreu-Rubinstein equilibrium, and hence a lean equilibrium (by Proposition 3.4), as follows.

First, we show that the sequence $\sigma$ is both 1-irreducible and 2-irreducible. We begin by arguing 1-irreducibility. Let $t_1, t_2 \in \{1, \ldots, k\}$ be two distinct values, and assume without loss of generality that $t_1 < t_2$. We show that $t_1$ and $t_2$ are 1-incompatible with respect to $\langle \sigma \rangle$. If $\sigma_1^{t_1} \neq \sigma_1^{t_2}$, then clearly $t_1$ and $t_2$ are 1-incompatible. In the case that $\sigma_1^{t_1} = \sigma_1^{t_2}$, by the definition of $\sigma$, there exists a minimum value $m \geq 1$ such that $\sigma_1^{t_2} \neq \langle \sigma \rangle_1^{t_2+m}$; observe that $\sigma_1^{t_1} = \langle \sigma \rangle_1^{t_1+m}$. For all $n$ with $0 \leq n < m$, we have $\langle \sigma \rangle_2^{t_1+n} = \langle \sigma \rangle_2^{t_2+n} = \sigma_1^{t_2}$, and so we have that $t_1$ and $t_2$ are 1-incompatible. We thus have that $\sigma$ is 1-irreducible; by an argument that is identical up to swapping the players, we also have that $\sigma$ is 2-irreducible.

We now argue that the pair $(M_1^\sigma, M_2^\sigma)$ is an Abreu-Rubinstein equilibrium with respect to $|R|$ and $||\delta||$. Observe that for these machines, we have $|R_1| = |R_2| = ||\delta_1|| = ||\delta_2|| = k$. We show by contradiction that there is no player 1 best response $N_1$ to $M_2^\sigma$ having $|R_1| < k$ or $||\delta_1|| < k$. Suppose that there is; then the payoff profile of $(N_1, M_2^\sigma)$ must be $(r_1(\sigma), r_2(\sigma))$, which is strictly enforceable, and by Proposition 4.1, it holds that $|P_1| < k$. By the 1-irreducibility of $\sigma$ and Theorem 5.2, it holds that $\equiv_1$ is equal to $\equiv_s$ and hence that $|P_1| = k$, contradicting that $|P_1| < k$. It can similarly be shown that there is no player 2 best response $N_2$ to $M_1^\sigma$ having $|R_2| < k$ or $||\delta_2|| < k$. We have thus argued that the pair $(M_1^\sigma, M_2^\sigma)$ is an Abreu-Rubinstein equilibrium with respect to $|R|$ and $||\delta||$.

In the Prisoner's Dilemma, it can similarly be shown that for constants $N_{CD}, N_{DC} \geq 1$, the finite action sequence $\sigma = N_{CD} \cdot (C,D), N_{DC} \cdot (D,C)$ is both 1-irreducible and 2-irreducible, and that when $\sigma$ is strictly enforceable, the pair $(M_1^\sigma, M_2^\sigma)$ is an Abreu-Rubinstein equilibrium with respect to both $|R|$ and $||\delta||$. More generally, let $G = (S_1, S_2, u_1, u_2)$ be a game, let $\beta_1 : \{1, \ldots, b\} \to S_1$ and $\beta_2 : \{1, \ldots, b\} \to S_2$ be injective mappings, and let $N_1, \ldots, N_b \geq 1$ be constants, with $b \geq 2$. By arguments similar to those given above, the finite action sequence $\sigma = N_1 \cdot (\beta_1(1), \beta_2(1)), \ldots, N_b \cdot (\beta_1(b), \beta_2(b))$ can be shown to be both 1-irreducible and 2-irreducible, and when $\sigma$ is strictly enforceable, the pair $(M_1^\sigma, M_2^\sigma)$ can be shown to be an Abreu-Rubinstein equilibrium with respect to $|R|$ and $||\delta||$. $\square$

We now introduce another technique for establishing lean equilibria. Let $\sigma$ be a finite action sequence. Define a *rotation* of $\sigma = \sigma^1 \ldots \sigma^k$ to be a length $k$ sequence of the form $\sigma^n \sigma^{n+1} \ldots \sigma^k \sigma^1 \sigma^2 \ldots \sigma^{n-1}$ for $n$ with $1 \leq n \leq k$. Let $i \in \{1, 2\}$ be one of the players and let $B \subseteq S_i$. We say that $\sigma$ is $(i, B)$-*rigid* if for every rotation $\rho$ of $\sigma$ and every $n$ with $1 \leq n < k$, when $\rho_i^1, \rho_i^{n+1} \in B$, it holds that $(u_j(\rho_1) + \cdots + u_j(\rho_n))/n \neq r_j(\sigma)$.

**Theorem 5.4** *Let $i \in \{1, 2\}$ and $B \subseteq S_i$. Let $\sigma$ be a strictly enforceable finite action sequence of length $k$, let $b$ be the number of elements $\sigma^n$ of $\sigma$ with $\sigma_i^n \in B$, and let $M_j$ be a $\sigma$-machine. If $\sigma$ is $(i, B)$-rigid, then for any machine $N_i$ with $|\{q \in P_i \mid \lambda_i(q) \in B\}| < b$ relative to $(N_i, M_j)$, the pair $(N_i, M_j)$ is not a Nash equilibrium.*

**Proof**. Since $M_j$ is a $\sigma$-machine, if $(s^t)$ is not equal to $\langle \sigma \rangle$, then $N_i$ is not a best response to $M_j$ and $(N_i, M_j)$ is not a Nash equilibrium, so we assume that $(s^t) = \langle \sigma \rangle$.

Consider the state pair sequence $(q^{1+dk})_{d \geq 1}$. By the finiteness of the state sets of the machines, some state pair must occur infinitely often in this sequence. Hence we can find time points $t_1, t_2$ of the form $1+dk$ with $t_1 < t_2$ such that $q^{t_1} = q^{t_2}$. The sequence $q^{t_1}, q^{t_1+1}, \ldots, q^{t_2}$ determines a cycle

$$C = q_i^{t_1} \xrightarrow{s_j^{t_1}} q_i^{t_1+1} \xrightarrow{s_j^{t_1+1}} \cdots q_i^{t_2}$$

in $N_i$. By our choice of $t_1, t_2$ and the assumption that $\langle s^t \rangle = \langle \sigma \rangle$, we have $r_j(C) = r_j(N_i, M_j^\sigma) = r_j(\sigma)$.

By the hypothesis on $B$ and $b$, among the sequence of states $q_i^{t_1}, q_i^{t_1+1}, \ldots, q_i^{t_1+(k-1)}$ there must be two indices $t' < t''$ with $q_i^{t'} = q_i^{t''}$ and $\lambda(q_i^{t'}) \in B$. The sequence of states $q_i^{t'}, q_i^{t'+1}, \ldots, q_i^{t''}$ determine a subcycle $C'$ of $C$ which, by $(i, B)$-rigidity, has $r_j(C') \neq r_j(C)$. We can view $C$ as the concatenation of the subcycle $C'$ with another subcycle $C''$. The value $r_j(C)$ is the convex combination of $r_j(C')$ and $r_j(C'')$; since $r_j(C') \neq r_j(C)$, we have $r_j(C'') \neq r_j(C)$. It must hold that one of the values $r_j(C'), r_j(C'')$ is strictly greater than $r_j(C)$. This implies that $(N_i, M_j)$ is not a Nash equilibrium, as player $j$ could strictly improve his payoff by deviating. $\square$

**Example 5.5** Consider, in the Prisoner's Dilemma, a strictly enforceable payoff $w$ that is the convex combination of $u(C, C)$ and $u(C, D)$. We can write $w = (N_{CC}/N)u(C, C) + (N_{CD}/N)u(C, D)$ where $N_{CC}, N_{CD}$ are integers, and $N = N_{CC} + N_{CD}$. Since the payoff $w$ is strictly enforceable, we have $N_{CC} > 0$. We can assume that $N_{CC}, N_{CD}$ do not share any prime factors, for if they do share one, we can divide both of them by the factor while preserving the value of $w$. Note that this assumption implies that $N_{CC}$ and $N$ do not share any prime factors.

Let $\sigma = N_{CC} \cdot (C, C), N_{CD} \cdot (C, D)$. We will show that $(M_1^\sigma, M_2^\sigma)$ is a lean equilibrium with respect to $|R|$ and $||\delta||$. Note that the first player could preserve payoff but reduce complexity via a machine with one state that always outputs $C$, a machine that has $|R| = 1$ and $||\delta|| = 2$. Hence this pair is *not* an Abreu-Rubinstein equilibrium with respect to $|R|$ when $N \geq 2$, nor with respect to $||\delta||$ when $N \geq 3$. Along these lines, observe that no two $\equiv_s$-equivalence classes (for $\langle \sigma \rangle$) are 1-incompatible, since in the sequence $\sigma$ player 1 always plays the same action.

We show that $\sigma$ is $(1, \{C\})$-rigid. We show this by contradiction; let $\rho$ be a rotation of $\sigma$ and let $n$ be such that $1 \leq n < N$ and $(u_2(\rho_1) + \cdots + u_2(\rho_n))/n = w$. This implies that for integers $n_{CC}, n_{CD}$ with $0 \leq n_{CC} \leq n, 0 \leq n_{CD} \leq n$, and $n_{CC} + n_{CD} = n$, we have $(n_{CC}/n)u(C, C) + (n_{CD}/n)u(C, D) = w$. Since $u(C, C) \neq u(C, D)$, we have $n_{CC}/n = N_{CC}/N$, implying that $N_{CC}n = n_{CC}N$. This implies that $N > 1$ divides $N_{CC}n$. Since $N$ and $N_{CC}$ do not share any prime factors, this implies that $N$ divides $n$, a contradiction to $n < N$. We have thus shown that $\sigma$ is $(1, \{C\})$-rigid.

Consider any machine $N_1$ with $|R_1| < N$ or $||\delta_1|| < N$. Such a machine must have $|P_1| < N$, and hence by Theorem 5.4, the pair $(N_1, M_2^\sigma)$ is not a Nash equilibrium. On the other hand, it is straightforward to verify that the sequence $\sigma$ is 2-irreducible. Thus, for any machine $N_2$ with $|R_2| < N$ or $||\delta_2|| < N$, we have $|P_2| < N$ and by Theorem 5.2, the machine $N_2$ is not a best response to $M_1^\sigma$. We conclude that the pair $(M_1^\sigma, M_2^\sigma)$ is a lean equilibrium with respect to $|R|$ and $||\delta||$. $\square$

**Example 5.6** Consider, in the Prisoner's Dilemma, a finite action sequence of the form $\sigma = k_{CD} \cdot (C, D), k_{DD} \cdot (D, D), k_{DC}(D, C)$, where $k_{CD}, k_{DD}, k_{DC} \geq 1$ and $\sigma$ is strongly enforceable. We use $k$ to denote the length $k_{CD} + k_{DD} + k_{DC}$ of $\sigma$. We show that the pair $(M_1^\sigma, M_2^\sigma)$ is a lean equilibrium with respect to $|R|$ and $||\delta||$.

Let $N_1$ be a machine for player 1. Suppose that $N_1$ is a best response to $M_2^\sigma$. Then the pair $(N_1, M_2^\sigma)$ produces the action sequence $\langle s^t \rangle = \langle \sigma \rangle$. We show that if $N_1$ has $|R_1| < k$ or $||\delta_1|| < k$, then the pair $(N_1, M_2^\sigma)$ is not a Nash equilibrium.

Let $Q_C$ denote the states of $N_1$ that output $C$, and let $Q_D$ denote the normal states of $N_1$ that output $D$. It is straightforward to verify that for any two distinct $t_1, t_2 \in \{1, \ldots, k_{CD}\}$, the classes $[t_1]_s, [t_2]_s$ are 1-incompatible, and hence, we have $|Q_C| \geq k_{CD}$ by Proposition 5.1. We next show that $\sigma$ is $(1, \{D\})$-rigid. Consider any rotation $\rho$ of $\sigma$ and a value $n$ with $1 \leq n < k$ and $\rho_1^1 = \rho_1^{n+1} = D$; in one of the sequences $\rho' = \rho_1 \ldots \rho_n, \rho'' = \rho_{n+1} \ldots \rho_k$, player 1 uses only the action $D$ and hence one of the values $r_2(\rho'), r_2(\rho'')$ is strictly below 0. On the other hand, $r_2(\sigma)$ is strictly above 0 and can be written as the convex combination of $r_2(\rho')$ and $r_2(\rho'')$, and so neither of $r_2(\rho'), r_2(\rho'')$ is equal to $r_2(\sigma)$, and we have that $\sigma$ is $(1, \{D\})$-rigid. Now suppose that $N_1$ has $|R_1| < k$ or $||\delta_1|| < k$. It follows that $|P_1| < k$; since $|Q_C| \geq k_{CD}$, this implies that $|Q_D| < k_{DD} + k_{DC}$. By Theorem 5.4, we have that $(N_1, M_2^\sigma)$ is not a Nash equilibrium.

In a similar way, it can be shown that for any best response $N_2$ to $M_1^\sigma$, if $N_2$ has $|R_2| < k$ or $||\delta_2|| < k$, then the pair $(M_1^\sigma, N_2)$ is not a Nash equilibrium. We thus have that $(M_1^\sigma, M_2^\sigma)$ is a lean equilibrium with respect to $|R|$ and $||\delta||$. □

So far, our discussion has focused on the complexity measures $|R|$ and $||\delta||$. We now turn our attention to the complexity measure $|Q|$.

**Example 5.7** As in the previous example, let $\sigma$ be a finite action sequence of the form $\sigma = k_{CD} \cdot (C, D), k_{DD} \cdot (D, D), k_{DC}(D, C)$ for the Prisoner's Dilemma, where $k_{CD}, k_{DD}, k_{DC} \geq 1$ and $\sigma$ is strongly enforceable; let $k$ denote the length $k_{CD} + k_{DD} + k_{DC}$ of $\sigma$. We give a pair of machines $(M_1, M_2)$ where each machine has $k$ states that is a lean equilibrium with respect to $|Q|$.

We define the machines $M_1, M_2$ as follows. Each machine has state set $Q_1 = Q_2 = \{1, \ldots, k\}$, initial states $q_1^1 = q_2^1 = 1$, and has output function defined by $\lambda_i(n) = \sigma_i^n$ for all $n \in Q_i$. For the states $n \in Q_i$ where $\sigma_j^n = D$, we define $\delta_i(n, C) = \delta_i(n, D) = n + 1$, where $k + 1$ is understood to represent the state 1. For the states $n \in Q_i$ where $\sigma_j^n = C$, we define $\delta_i(n, C) = n + 1$ and $\delta_i(n, D) = q_i^*$ where $q_i^*$ is the first state where $\sigma_j^n = C$, that is, $q_1^* = k_{CD} + k_{DD} + 1$ and $q_2^* = 1$. The states $q_i^*$ can be thought of as "internal threat" states. This construction is similar to that of [1, Page 1276, Case B].

We now observe that in each machine $M_i$, the simple cycle $C_i$ that maximizes payoff to the other player $j$ is the cycle naturally corresponding to $\sigma$, that is, $1 \xrightarrow{\sigma_j^1} 2 \xrightarrow{\sigma_j^2} \cdots k \xrightarrow{\sigma_j^k} 1$; this cycle has $r_j(C_i) = r_j(\sigma)$. This is clearly the unique payoff-maximizing simple cycle of length $k$. Also, in all shorter simple cycles, player $i$ only defects, yielding player $j$ a payoff strictly less than 0.

Let $N_i$ be a best response to $M_j$. By the observation in the previous paragraph, the sequence $(q_j^t)$ must, after some finite amount of time, be equal to the sequence $1, \ldots, k$ repeated infinitely. It is hence possible to modify the machine $N_i$, by changing its initial state to a state that is played against the state $q_j^1 = 1$ in the mentioned infinite repetition, to obtain a machine $N_i'$ that, along with $M_j$, generates the sequence $\langle \sigma \rangle$, and has the same number of states as $N_i$. Now, if $N_i'$ and $N_i$ have strictly fewer than $k$ states, then against $M_j$ they have strictly fewer than $k$ played states, and then by arguing as in Example 5.6, the pair $(N_i', M_j)$ is not a Nash equilibrium, from which it follows that the pair $(N_i, M_j)$ is not a Nash equilibrium. We conclude that $(M_1, M_2)$ is a lean equilibrium with respect to $|Q|$. □

We now establish a theorem that will help us to establish lean equilibrium results with respect to $|Q|$. Let us say that $\sigma$ is *i-foolable* if there exists a rotation $\rho = \rho^1 \ldots \rho^k$ of $\sigma$ and an action $s' \in S_j$ such that for all $n$ with $1 \leq n \leq k$, it holds that $r_j(\rho^n \rho^{n+1} \ldots \rho^{k-1} \rho') > r_j(\sigma)$, where $\rho'$ is the pair with player $i$ action equal to $\rho_i^k$ and player $j$ action equal to $s'$.

**Theorem 5.8** *Let $\sigma$ be a strictly enforceable finite action sequence, and let $M_j$ be a $\sigma$-machine. If $\sigma$ is i-foolable (via $\rho$), and $N_i$ is a machine such that in $(N_i, M_j)$ it holds that $P_i = Q_i$ (that is, all states in $N_i$ are played), then $(N_i, M_j)$ is not a Nash equilibrium.*

**Proof**. If $N_i$ is not a best response to $M_j$, we are done, so we assume that $N_i$ is a best response to $M_j$, in which case we have $(s^t) = \langle \sigma \rangle$. Let $q_i$ be any state of $Q_i$. We claim that there is a path $P$ in $N_i$ from $q_i$ to a state $q_i' \in Q_i$ such that $r_j(P) > r_j(\sigma)$. This suffices, since it implies that $M_j$ is not a best response to $N_i$.

We reason as follows. By hypothesis, the state $q_i$ is played and hence there exists $t \geq 1$ with $q_i = q_i^t$. Since $(s^t) = \langle \sigma \rangle$, there exists $n$ with $1 \leq n \leq k$ where $\rho^n \rho^{n+1} \ldots \rho^k = s^t s^{t+1} \ldots s^{t+(k-n)}$. The desired path $P$ starts at $q_i$ and has actions $\rho_j^n \ldots \rho_j^{k-1} s'$, where $s' \in S_j$ is the action from the definition of $i$-foolable; by that definition, we have $r_j(P) > r_j(\sigma)$. □

**Example 5.9** We return to the class of sequences considered in our first example, Example 5.3. Let $N_C, N_D \geq 1$ be constants, and consider the finite action sequence $\sigma = N_C \cdot (C,C), N_D \cdot (D,D)$ in the Prisoner's Dilemma; the sequence $\sigma$ is strictly enforceable, and has length $k = N_C + N_D$. We show that the pair $(M_1^\sigma, M_2^\sigma)$ is a lean equilibrium with respect to $|Q|$. Note that this pair is *not* an Abreu-Rubinstein equilibrium with respect to $|Q|$, since each of the machines has a threat state that is never played, and hence each of the machines could be simplified without sacrificing payoff by removing this threat state.

To show that the described pair is a lean equilibrium, we show that, for each player $i$, when $N_i$ is a best response to $M_j^\sigma$ with $k$ or fewer states, the pair $(N_i, M_j^\sigma)$ is not a Nash equilibrium. By the argumentation in Example 5.3, any such best response $N_i$ must have at least $k$ played states. Hence in such a best response $N_i$, all states are played; by Theorem 5.8, it thus suffices to show that the sequence $\sigma$ is $i$-foolable. It is straightforward to verify that $\sigma$ is $i$-foolable via the rotation $\rho = N_D \cdot (D,D), N_C \cdot (C,C)$ and the action $D \in S_j$. □

**Example 5.10** We reconsider the sequences treated in Example 5.5. Let $\sigma = N_{CC} \cdot (C,C), N_{CD} \cdot (C,D)$, where $N_{CC} > 0$ and $N_{CC}, N_{CD}$ do not share any prime factors. We show that the pair $(M_1^\sigma, M_2^\sigma)$ is a lean equilibrium with respect to $|Q|$. This pair is not an Abreu-Rubinstein equilibrium with respect to $|Q|$: the first machine could be simplified to a machine that only outputs $C$ without giving up payoff, and the second machine could eliminate its threat state without giving up payoff.

We show that the sequence $\sigma$ is both 1-foolable and 2-foolable. We have 1-foolability by the rotation $(N_{CC} - 1) \cdot (C,C), N_{CD} \cdot (C,D), (C,C)$ and the action $D \in S_2$, and we have 2-foolability by the rotation $N_{CD} \cdot (C,D), N_{CC} \cdot (C,C)$ and the action $D \in S_1$.

We can now argue that the pair $(M_1^\sigma, M_2^\sigma)$ is a lean equilibrium with respect to $|Q|$. The structure of the argument is similar to that of the previous example. Consider a player $i$ and a best response $N_i$ to $M_j^\sigma$ with $k$ or fewer states. It is shown in Example 5.5 that if $N_i$ has strictly fewer than $k$ played states, then $(N_i, M_j^\sigma)$ is not a Nash equilibrium. In the case that $N_i$ has exactly $k$ played states, all of its states are played and then $(N_i, M_j^\sigma)$ is not a Nash equilibrium by Theorem 5.8. □

The results in the last three examples demonstrate different types of payoffs that are sustainable by lean equilibria with respect to $|Q|$ in the repeated Prisoner's Dilemma. In particular, Example 5.7 shows that any strictly enforceable payoff profile in the interior of the convex hull of the points $u(C,D), u(D,D)$, and $u(D,C)$, is a payoff attainable by such a lean equilibrium. These results can be contrasted strongly with the results of Abreu and Rubinstein [1, Section 5] showing that the payoffs of Abreu-Rubinstein equilibria in this context are the strictly enforceable payoffs that are convex combinations of the diagonals, that is, convex combinations of $u(C,C)$ and $u(D,D)$ and convex combinations of $u(C,D)$ and $u(D,C)$.

# 6 Structure of lean equilibria

In this section, we present results describing the structure of lean equilibria in machine games $G_m = (\mathcal{M}_1, \mathcal{M}_2, r_1, r_2)$ with respect to the complexity measures $|R|$ and $||\delta||$. Our first result demonstrates that the sequence $(q^t)$ begins with a sequence of state pairs where each state is used only once, followed by a state pair where each state is used infinitely often.

**Lemma 6.1** *(1-$\infty$ Lemma) Suppose that $(M_1, M_2)$ is a lean equilibrium of $G_m$ with respect to one of the complexity measures $|R|$, $||\delta||$ having a strictly enforceable payoff profile. Let $u \geq 1$ be the minimum value such that one of the states $q_1^u$, $q_2^u$ is used later in the sequence $(q^t)$, that is, such that there exists $i \in \{1, 2\}$ such that $q_i^u \in \{q_i^t \mid t > u\}$. Then, for each $i \in \{1, 2\}$, the state $q_i^u$ appears infinitely often in the sequence $(q_i^t)$.*

**Proof.** We prove this by contradiction. Suppose one or both of the states $q_1^u$, $q_2^u$ appears finitely often in the respective sequence $(q_i^t)$. Let $T_i(q)$ denote the set $\{t \mid q = q_i^t\}$, that is, the points in time where player $i$ plays state $q$. Observe that $T_1(q_1^u) \cap T_2(q_2^u) = \{u\}$, for if this intersection contains two distinct elements, there must be infinitely many points $t$ such that $q^t = q^u$.

We claim that the players can be labelled as $i'$, $i''$ in such a way that: (1) $T_{i'}(q_{i'}^u)$ is finite, and (2) there exists $v \in T_{i''}(q_{i''}^u)$ such that $v > \max T_{i'}(q_{i'}^u)$.

We establish this claim as follows. If both sets $T_1(q_1^u), T_2(q_2^u)$ are finite, set $v = \max(T_1(q_1^u) \cup T_2(q_2^u))$ and let $i''$ be the unique element in $\{1, 2\}$ such that $v \in T_{i''}(q_{i''}^u)$. If one of the sets $T_1(q_1^u), T_2(q_2^u)$ is finite and the other is infinite, let $i'$ be the player in $\{1, 2\}$ such that $T_{i'}(q_{i'}^u)$ is finite; since $T_{i''}(q_{i''}^u)$ is infinite, it is possible to select a value $v$ satisfying condition (2).

For the sake of notation, we now assume that $i' = 1$ and $i'' = 2$. We want to show that $(M_1, M_2)$ is not a lean equilibrium. If $(M_1, M_2)$ is not a Nash equilibrium, we are done, so we assume that it is. Starting from $M_1$, we define a new machine $M_1'$ as follows. We set $\delta_1'(q_1^{u-1}, s_2^{u-1}) = q_1^v$, or, in the case that $u = 1$, we set the initial state of $M_1'$ to be $q_1^v$. We then have that

$$q_1^1 \xrightarrow{s_2^1} q_1^2 \xrightarrow{s_2^2} \cdots q_1^{u-1} \xrightarrow{s_2^{u-1}} q_1^v$$

is a path in $M_1'$. We modify $M_1'$ so that, other than the transitions in this path, there are no transitions to the states $q_1^v, \ldots, q_1^{u-1}$; we reroute the transitions to these states to a threat state. We also eliminate the state $q_1^u$, rerouting the transitions to it to a threat state. The state sequence induced by the machine pair $(M_1', M_2)$ is $q^1, q^2, \ldots, q^{u-1}, q^v, q^{v+1}, q^{v+2}, \ldots$ and hence the payoffs to each of the two players is the same as in $(M_1, M_2)$.

We show that $(M_1', M_2)$ is a Nash equilibrium by arguing that player 2 cannot obtain a strictly better payoff. Let $C$ be any cycle in $M_1'$. None of the states $q_1^1, \ldots, q_1^u$ can appear in $C$; since all modified transitions involved these states, the cycle $C$ is also a cycle of $M_1$. As $M_2$ was a best response to $M_1$, it is a best response to $M_1'$.

We now need only argue that $M_1'$ is simpler than $M_1$. In $(M_1, M_2)$, the state $q_1^u$ is a played state of $M_1$, so by Proposition 4.1, we have $|R_1'| < |R_1|$. Also, the state $q_1^u$ contributes at least 1 to the value $||\delta_1||$, a contribution not present in the calculation of $||\delta_1'||$, so $||\delta_1'|| < ||\delta_1||$. $\square$

**Lemma 6.2** *Suppose that $(M_1, M_2)$ is a lean equilibrium of $G_m$ with respect to one of the complexity measures $|R|$, $||\delta||$ having a strictly enforceable payoff profile, and suppose that $\equiv_i$ is contained in $\equiv_s$ (for some $i \in \{1, 2\}$). Then, the equivalence relations $\equiv_s$, $\equiv_i$ are equal.*

**Proof.** We prove this by contradiction. Suppose there are two values $t'' < t'$ such that $t'' \equiv_s t'$ but $t'' \not\equiv_i t'$. Without loss of generality, we assume that $i = 1$, and so we have $q_1^{t''} \neq q_1^{t'}$. We want to show that $(M_1, M_2)$ is not a lean equilibrium; if $(M_1, M_2)$ is not a Nash equilibrium, we are done, so we assume that it is.

Define $M_1'$ to be the machine equal to $M_1$, but where the state $q_1^{t''}$ is eliminated and all transitions to $q_1^{t''}$ from states in $R_1 \setminus \{q_1^{t''}\}$ are changed to transitions to $q_1^{t'}$.

The machine $M_1'$ is simpler than $M_1$ with respect to the complexity measure $|R|$, as it has one fewer played state than $M_1$ (see Proposition 4.1). It is also simpler than $M_1$ with respect to $||\delta||$: the number of normal transitions out of states in $R_1'$ is equal to that of $R_1$, but the state $q_1^{t''}$ (in $M_1$) has at least one normal transition, as it is played in $(M_1, M_2)$.

We claim that $(M_1', M_2)$ is a Nash equilibrium of $G_m$, which suffices. Define $f([t]_s)$ as $\{q_1^u \in Q_1 \mid u \in [t]_s\}$. Let $(\hat{q}^t)$, $(\hat{s}^t)$ be the sequences induced by the machines $(M_1', M_2)$. We prove by induction that for all $t \geq 1$, it holds that $\hat{q}_1^t \in f([t])$, $\hat{q}_2^t = q_2^t$, and $\hat{s}^t = s^t$. The base case is clear, so assume that the claim holds on $t \geq 1$. Then $\hat{q}_1^t = q_1^{t'}$ for some $t' \equiv_s t$. As $\hat{s}_2^t = s_2^t = s_2^{t'}$, we have $\hat{q}_1^{t+1} = \delta_1(\hat{q}_1^t, \hat{s}_2^t) = \delta_1(q_1^{t'}, s_2^{t'}) = q_1^{t'+1}$. This implies that $\hat{q}_1^{t+1} \in f([t'+1]_s) = f([t+1]_s)$. As $\lambda_1(\hat{q}_1^t) = \lambda_1(q_1^{t'}) = \lambda_1(q_1^t) = s_1^t$, we have $\hat{q}_2^{t+1} = q_2^{t+1}$ and hence $\hat{s}^{t+1} = s^{t+1}$.

Thus, to show that $(M_1', M_2)$ is a Nash equilibrium of $G_m$, it suffices to show that $M_2$ cannot deviate to obtain a strictly higher payoff. Suppose that $p_1 \stackrel{a_1}{\to} p_2 \stackrel{a_2}{\to} \cdots p_m \stackrel{a_m}{\to} p_1$ is a $M_1'$-cycle $C$ giving $M_2$ a payoff $r > r_2(M_1, M_2) = r_2(M_1', M_2)$. By the choice of the values $t'', t'$, there is a path $P$ in $M_1$ from $q_1^{t''}$ to $q_1^{t'}$ whose $M_2$-payoff is $r_2(M_1, M_2)$. For each state-action pair $(p_j, a_j)$ in the cycle $C$ with $\delta_1(p_j, a_j) = q_1^{t''}$ (and hence $\delta_1'(p_j, a_j) = q_1^{t'}$), replace $p_{j+1} = q_1^{t'}$ with the path $P$. In this way, we obtain a $M_1$-cycle whose payoff is also strictly greater than $r_2(M_1, M_2) = r_2(M_1', M_2)$, contradicting that $(M_1, M_2)$ is a Nash equilibrium of $G_m$. □

With respect to a machine pair $(M_1, M_2)$, we let $P_i$ denote the set of played states of $M_i$.

**Lemma 6.3** *Suppose that $(M_1, M_2)$ has a strictly enforceable payoff profile. If $(M_1, M_2)$ is a lean equilibrium with respect to $|R|$, then $|R_1| = |R_2| = |P_1| = |P_2|$, and if $(M_1, M_2)$ is a lean equilibrium with respect to $||\delta||$, then $||\delta_1|| = ||\delta_2|| = |P_1| = |P_2|$.*

**Proof.** We claim that, starting from a Nash equilibrium $(M_1, M_2)$ with a strictly enforceable payoff profile, player $j$ has a best response $M_j'$ to $M_i$ where $(M_j', M_i)$ is a Nash equilibrium and such that $|R_j| \leq |P_i|$ and $||\delta_j|| \leq |P_i|$. This implies, in the case of a lean equlibrium with respect to $|R|$, that $|R_2| \leq |P_1| \leq |R_1| \leq |P_2| \leq |R_2|$, and similarly, in the case of a lean equilibrium with respect to $||\delta||$, that $|P_2| \leq ||\delta_2|| \leq |P_1| \leq ||\delta_1|| \leq |P_2|$. The claim is argued as follows. The payoffs $r_1(M_1, M_2), r_2(M_1, M_2)$ are equal to the payoffs to the two players of a cycle $C$ in machine $M_i$. The cycle $C$ is not necessarily a simple cycle, but if it is not simple, it can be viewed as the concatenation of two shorter cycles. Each of the two shorter cycles must give the same payoff to player $j$ (otherwise $(M_1, M_2)$ would not be a Nash equilibrium, as player $j$ could profitably deviate). We choose the cycle out of the two shorter cycles that gives the higher payoff to player $i$. We then iterate this process until we obtain a simple cycle $C'$. The simple cycle has $r_j(C') = r_j(C)$ and $r_i(C') \geq r_i(C)$. It is possible to implement a player $j$ machine $M_j'$ that repeatedly walks the cycle $C'$ in $M_i$ satisfying the stated inequalities: this is done by taking a machine that simply walks a shortest path from the initial state of $M_i$ to a state in the cycle $C'$, and then repeatedly walks the cycle $C'$. The pair $(M_j', M_i)$ is a Nash equilibrium: the machine $M_j'$ obtains the same payoff as the machine $M_j'$, and the machine $M_i$ could only profitably deviate by playing a threat state; but since his payoff is greater than or equal to his payoff in $(M_j, M_i)$, this is not beneficial as his payoff is strictly above his minmax payoff. □

We now present our main structure theorem. This theorem not only describes the structure of machines at lean equilibrium with respect to the measure $||\delta||$, but shows that their structure can be derived solely from the equivalence relation $\equiv_s$, and hence just from the action sequence $(s^t)$; this implies that a third-party observer that only views the resulting action sequence can infer the structure of the machines. In order to give the statement, we introduce the following notion. Define a *rho-machine* to be a machine $M$ where each normal state $q$ reachable from the initial state has exactly one outgoing transition to a normal state;

denote this successor state by $s(q)$. We call the set of states occurring finitely often in the sequence $q^1$, $s(q^1)$, $s^2(q^1)$, ... the *tail* of the machine, and the other states (those occurring infinitely often) the *head* of the machine.

**Theorem 6.4** *Suppose that $(M_1, M_2)$ is a lean equilibrium of $G_m$ with respect to the complexity measure $||\delta||$ having a strictly enforceable payoff profile. Then, the equivalence relations $\equiv_s$, $\equiv_q$, $\equiv_1$, $\equiv_2$ are all equal, and for each $i \in \{1, 2\}$, the machine $M_i$ is a rho-machine having exactly one state $q_i([t]_s)$ that is played at all time points $[t]_s$ for each equivalence class $[t]_s$ of $\equiv_s$, whose structure is given by $\delta_i(q_i([t]_s), s_j^t) = q_i([t+1]_s)$ for all $t \geq 1$.*

**Proof**. By Lemma 6.3, in each of the machines $M_1$, $M_2$, each played state has exactly one outgoing normal transition which is to another played state. Thus, each of the machines is a rho-machine. By Lemma 6.1, the machines have the same tail size and the same head size, and thus the equivalence relations $\equiv_1$ and $\equiv_2$ are the same, from which it follows (by definition of $\equiv_q$) that the equivalence relations $\equiv_1$, $\equiv_2$, and $\equiv_q$ are all the same. Since $\equiv_q$ is always contained in $\equiv_s$, we can invoke Lemma 6.2 to obtain that $\equiv_1$ and $\equiv_2$ are each equal to $\equiv_s$, and we have that all four of the equivalence relations are equal.

For each $i \in \{1, 2\}$, by the equivalence of $\equiv_i$ and $\equiv_s$, the machine $M_i$ has one played state for each equivalence class of $\equiv_s$. As already noted, each played state has exactly one outgoing normal transition which is to another played state, and so the machine must be a rho-machine with the described structure. □

## 7 Discussion

We introduced and studied the notion of lean equilibrium, a particular form of Nash equilibrium where strategies cannot be further simplified according to a cautious simplification procedure: a player simplifies only if post-simplification, the strategy vector will be a Nash equilibrium. It is possible to consider similar equilibrium notions relative to even more cautious simplification procedures: for instance, a player might anticipate simplifications of other players, and only want to simplify if, in addition to preserving Nash equilibrium, he will not lose payoff if other players simplify following his simplification. A variant of this idea would have a player simplifying if he will not lose payoff in the case that other players change best response following his simplification. We leave the investigation of these equilibrium notions to future work. The broad research direction that we hope to have identified in the present work is that of investigating notions of equilibria where players prefer simple strategies, but where the desire for simplicity is not wired directly into the players' payoffs.

**Acknowledgements.** The author thanks Rani Spiegler and Olivier Gossner for their useful comments and pointers. The author has also benefited from feedback from participants in talks given in the 9th Conference on Logic and the Foundations of Game and Decision Theory (LOFT 2010), École Polytechnique (Department of Economics), and the Barcelona Jocs seminar. The author is indebted to the two anonymous IJGT reviewers for their careful consideration of this paper. The author is supported by the Spanish program "Ramon y Cajal" and MICINN grant TIN2010-20967-C04-02.

## References

[1] Dilip Abreu and Ariel Rubinstein. The structure of nash equilibrium in repeated games with finite automata. *Econometrica*, 56(6):1259–1281, 1988.

[2] Luca Anderlini. Some notes on church's thesis and the theory of games. *Theory and Decision*, 29:19–52, 1990.

[3] R. Aumann. Perspectives on bounded rationality. In *Proceedings of 4th International Conference on Theoretical Aspects of Reasoning about Knowledge*, pages 108–117, 1992.

[4] J. Banks and R. Sundaram. Repeated games, finite automata and complexity. *Games and Economic Behavior*, 2:97–117, 1990.

[5] Elchanan Ben-Porath. The complexity of computing a best response automaton in repeated games with mixed strategies. *Games and Economic Behavior*, 2:1–12, 1990.

[6] Elchanan Ben-Porath. Repeated games with finite automata. *Journal of Economic Theory*, 59:17–32, 1993.

[7] E. Ben-Sasson, A. Kalai, and E. Kalai. An approach to bounded rationality. In *Advances in Neural Information Processing Systems 19 (Proc. of NIPS)*, 2006.

[8] K. Binmore. Modeling rational players I. *Economics and Philosophy*, 3:179–214, 1987.

[9] K. Binmore. Modeling rational players II. *Economics and Philosophy*, 4:9–55, 1988.

[10] David Canning. Rationality, computability, and nash equilibrium. *Econometrica*, 60, 1988.

[11] L. Fortnow. Program equilibria and discounted computation time. In *Proceedings of the 12th Conference on Theoretical Aspects of Rationality and Knowledge*, pages 128–133, 2009.

[12] Lance Fortnow and Rahul Santhanam. Bounding rationality by discounting time. *CoRR*, abs/0911.3162, 2009.

[13] I. Gilboa. The complexity of computing best response automata in repeated games. *J. Econ. Theory*, 45:342–352, 1988.

[14] Olivier Gossner. Repeated games played by cryptographically sophisticated players. Working paper, 1999.

[15] Joseph Y. Halpern and Rafael Pass. Game theory with costly computation. *CoRR*, abs/0809.0024, 2008.

[16] J. V. Howard. Cooperation in the prisoner's dilemma. *Theory and Decision*, 24(3):203–213, 1988.

[17] E. Kalai. Bounded rationality and strategic complexity in repeated games. In *Game Theory and Applications*, pages 131–157. Academic Press, 1990.

[18] E. Kalai and W. Stanford. Finite rationality and interpersonal complexity in repeated games. *Econometrica*, 56:397–410, 1988.

[19] N. Megiddo and A. Wigderson. On play by means of computing machines. In *Proceedings of the 1986 Conference on theoretical aspects of reasoning about knowledge*, pages 259–274, 1986.

[20] A. Neyman. Bounded complexity justifies cooperation in the finitely repeated prisoners' dilemma. *Economics Letters*, 19:227–229, 1985.

[21] A. Neyman. Cooperation, repetition, and automata. In *Cooperation: Game-Theoretic Approaches, S. Hart, A. Mas Colell, eds., NATO ASI Series F, Vol. 155*, pages 233–255. Springer-Verlag, 1997.

[22] Martin Osborne and Ariel Rubinstein. *A Course in Game Theory*. MIT Press, 1994.

[23] C. H. Papadimitriou. On players with a bounded number of states. *Games and Economic Behavior*, 1992.

[24] Christos H. Papadimitriou and Mihalis Yannakakis. On complexity as bounded rationality (extended abstract). In *STOC 1994*, pages 726–733, 1994.

[25] M. Piccione and A. Rubinstein. Finite automata play a repeated extensive game. *Journal of Economic Theory*, 61:160–168, 1993.

[26] A. Rubinstein. Finite automata play the repeated prisoner's dilemma. *Journal of Economic Theory*, 38:83–96, 1986.

[27] Ariel Rubinstein. *Modeling Bounded Rationality*. MIT Press, 1998.

[28] H. Simon. *The sciences of the artificial*. MIT Press, 1969.

[29] R. Spiegler. Simplicity of beliefs and delay tactics in a concession game. *Games and Economic Behavior*, 47:200–220, 2004.

[30] R. Spiegler. Testing threats in repeated games. *Journal of Economic Theory*, 121:214–235, 2005.

[31] M. Tennenholtz. Program equilibrium. *Games and Economic Behavior*, 49:363–373, 2004.